\documentclass[11pt]{article}

\usepackage{amsmath,amssymb,epsfig,cite,color,verbatim,array,bm,amsfonts,enumerate,enumitem}
\usepackage{graphics,float}
\usepackage[colorlinks=true,citecolor=ForestGreen,urlcolor=black,linkcolor=blue]{hyperref}
\usepackage{blindtext}
\usepackage{microtype}
\usepackage[latin9]{inputenc}
\usepackage{authblk,booktabs,hhline}
\usepackage[table,dvipsnames]{xcolor}

\definecolor{light-gray}{gray}{0.9}

\numberwithin{equation}{section}
\allowdisplaybreaks
\title{Symplectic quantization of multi-field Generalized Proca electrodynamics}

\author[1,2]{Ver\'onica Errasti D\'iez\footnote{vero.erdi@origins-cluster.de}}
\author[3,4]{Marina Krstic Marinkovic\footnote{marinama@ethz.ch}}

\affil[1]{ {\small {\it Excellence Cluster ORIGINS, Boltzmannstra{\ss}e 2, D-85748 Garching bei M\"unchen, Germany}}}
\affil[2]{ {\small {\it Universit{\"a}ts-Sternwarte, Fakult{\"a}t f\"ur Physik, Ludwig-Maximilians-Universit\"at M\"unchen, Scheinerstra{\ss}e 1, D-81679 M\"unchen, Germany}}}
\affil[3]{ {\small {\it Institute for Theoretical  Physics,  ETH  Z\"urich, Wolfgang-Pauli-Stra{\ss}e 27,
CH-8093  Z\"urich,  Switzerland}}}
\affil[4]{ {\small {\it Arnold-Sommerfeld-Center for Theoretical Physics, Ludwig-Maximilians-Universit\"at,
Theresienstr. 37, 80333 M\"unchen, Germany}}}

\date{}

\begin{document}

\maketitle

\begin{abstract}
We explicitly carry out the symplectic quantization of a family of multi-field Generalized Proca (GP) electrodynamics theories.
In the process, we provide an independent derivation of the so-called secondary constraint enforcing relations ---
consistency conditions that significantly restrict the allowed interactions in multi-field settings already at the classical level.
Additionally, we unveil the existence of quantum consistency conditions,
which apply in both single- and multi-field GP scenarios.
Our newly found conditions imply that not all classically well-defined (multi-)GP theories are amenable to quantization.
The extension of our results to the most general multi-GP class is conceptually straightforward, albeit algebraically cumbersome.
\end{abstract}

\section{Introduction}
\label{sec:intro}

Quantum electrodynamics (QED) is the commonly employed
relativistic quantum field theory of the electromagnetic force.
Even so, generalizations of QED are relevant in many branches of physics, including condensed matter, cosmology, optics, particle physics and string theory, e.g.~\cite{Tajmar,DeFelice:2016yws,Gorelik,Ellis:2017edi,Fradkin}.
Here, we derive the partition function of some recently proposed extensions of QED,
which comprise an arbitrary number of massive photons with derivative (self-)interactions.
The renowned quantization procedure put forward by Dirac, repeatedly refined and extended since its inception,
would be the standard approach to achieve this goal.
However, owing to the noteworthy difficulty of its implementation in our targeted class of theories,
we resort to the distinct yet physically equivalent symplectic quantization methodology instead.

It was almost 160 years ago that Maxwell laid the foundations of classical electromagnetism~\cite{Maxwell}.
Viewed as a field theory, this describes an abelian massless vector field and its linear interactions with sources.
The quantization of Maxwell's theory took several decades,
earned some of its key developers a Nobel Prize in 1965
and yielded what arguably remains the most successful theory to date: QED.
For a historical review, we refer the reader to~\cite{Bovy}.

As is well-known and was nicely recapped in~\cite{Bovy}, early attempts at quantizing electromagnetism
met with a divergent self-energy for any static point particle, such as the electron, placed in an electromagnetic field.
In order to overcome this problem, two fundamentally different modifications to Maxwell's theory were introduced.
In 1934, Born and Infeld proposed a certain non-linear extension,
which is gauge-invariant and contains a single free parameter~\cite{Born:1934gh}.
On the other hand, in the period of 1936-1938, Proca constructed a massive version of Maxwell's electrodynamics~\cite{Proca1,Proca2}, which explicitly breaks the gauge symmetry.
The Born-Infeld (BI) model is a concrete realization of what ultimately became a large class of theories~\cite{Schwinger1,Schwinger2,Pleb},
collectively known as non-linear electrodynamics (NLE).
For an excellent recent review of NLE, see~\cite{Sorokin:2021tge}.
Contrariwise, Proca electrodynamics rapidly became and remains cornerstone to optics in its original form~\cite{Bloem,Partanen_2017,Garc_a_2017,Exps}.

It is only comparatively recently, in 2014, that classical, non-linear extensions of Proca's massive electromagnetism,
containing {\it derivative} self-interactions of the vector field, were put forward~\cite{Tasinato:2014eka,Heisenberg:2014rta}.
These conform a vast class of theories, usually referred to as Generalized Proca (GP) or Vector Galileon.
The axiomatization and non-trivial extension to multiple fields of GP electrodynamics was carried out in~\cite{ErrastiDiez:2019trb,ErrastiDiez:2019ttn}.
It is this class of theories, (multi-)GP electrodynamics, whose quantization we shall focus on.
For the ease of the reader, we note that GP can be understood
as the massive counterpart to the more familiar class of NLE theories\footnote{Our lightning review
of extensions of classical electromagnetism is limited to
theories described by first-order Lagrangian densities.
Higher-order generalizations are of course possible.
On the massless side, the most renowned example is that of
Podolsky electrodynamics~\cite{Podolsky:1942zz,Podolsky:1944zz}.
On the massive side, there exists a single proposal so far: Proca-Nuevo~\cite{deRham:2020yet,deRham:2021mqq},
which can also be extended through some GP interaction terms~\cite{deRham:2021efp}.},
see table \ref{table:EMtable}.
We highlight the relevance of the multi-field settings:
they allow for non-abelian augmentations of GP, upon imposing the desired group structure in the field space.

\begin{table}[!t]  
\renewcommand*{\arraystretch}{1.9}
\centering
\begin{tabular}{|>{\centering\arraybackslash}m{3cm}|>{\centering\arraybackslash}m{6cm}|>{\centering\arraybackslash}m{6cm}|}
\hhline{~|-|-|}
\multicolumn{1}{c|}{} & \cellcolor{light-gray}  {\bf Linear} &  \cellcolor{light-gray}  {\bf Non-linear} \\ \hline
\cellcolor{light-gray}  {\bf Massless} & Maxwell   & Non-linear electrodynamics (NLE)   \\ \hline
\cellcolor{light-gray}  {\bf Massive} & Proca & Generalized Proca (GP)  \\ \hline
\end{tabular}
\caption{Classification of single-field electromagnetic theories,  whose Lagrangian density is manifestly first-order.
Both NLE and GP stand for populous classes of such theories.
In this work, we shall consider the non-trivial multi-field extension of the GP class, constructed in~\cite{ErrastiDiez:2019trb,ErrastiDiez:2019ttn}.}
\label{table:EMtable}
\end{table}

To date, the phenomenology of (multi-field) GP theories has been fruitfully exploited in the context of cosmology, after their coupling to gravity.
Remarkable studies in this regard include the viable late-time acceleration scenarios in~\cite{DeFelice:2016yws}, the alleviation of the so-called $H_0$ tension in~\cite{Heisenberg:2020xak} and the primordial inflationary solutions in~\cite{Garnica:2021fuu}.
The most stringent empirical constraints on the free parameters of GP theories in a curved background follow from the measured propagation of gravitational waves~\cite{LIGOScientific:2017vwq}.
However, the vast free parameter space of GP theories is far from being ruled out by this
and other observations.
Therefore, our subsequent quantization of (multi-)GP theories in flat spacetime should be regarded as an important non-trivial
step towards the promising extension of the above investigations to the quantum realm.

Additionally, we advance the premise that quantum GP theories also show a noteworthy, although virtually unexplored, potential already in flat spacetime.
On the one hand, they allow for a towering generalization of the 
physical equivalence between
Maxwell electrodynamics in certain media and Proca electrodynamics in the vacuum,
which is the theoretical basis of the
prevalence of Proca's theory in optics.
In this regard, quantum GP can 
effectively describe light propagation
in a much wider set of media than Proca theory. 
Free parameters in GP will then need to be fine-tuned to match the dielectric constant of the material of interest~\cite{Mikki}.
On the other hand, the imminent, first-ever experimental probes of the non-linear regime of QED
--- most prominently by PVLAS~\cite{Ejlli:2020yhk} and LUXE~\cite{Abramowicz:2021zja} ---
necessitate strong theoretical foundations to
model the forthcoming observations.
In this context, GP goes hand in hand with 
NLE,
the chief constraint on its free parameters coming from the upper bound on the mass of the photon~\cite{Lakes,Chibi}.

All the theories mentioned so far are singular or constrained.
Further examples are non-abelian gauge field theories, gravitational theories and supersymmetric theories.
The systematic study of such systems was initiated by Dirac in 1950~\cite{Dirac:1950pj},
whose work was promptly and abundantly followed upon~\cite{Anderson:1951ta,BergGold,Gotayetal,Batlleetal,LeeWald},
including recent advancements~\cite{Vitagliano:2009fn,Pitts:2013uxa,PrMarRomRoy,Crisostomi:2017aim,ErrastiDiez:2020dux}.
In particular, the path integral formulation of Dirac's canonical quantization procedure
has been known for over four decades~\cite{Fadpath,Senjanovic:1976br}.

The formalism instituted by Dirac is ubiquitous but not unique.
In the present manuscript, we will employ the distinct quantization scheme introduced by Faddeev and Jackiw in 1988~\cite{Faddeev:1988qp}.
This method is conceptually simpler and, for some theories, it is algebraically easier to implement as well. 
The main reason for the conceptual simplicity lies in the fact that
Faddeev and Jackiw's approach does {\it not} require to classify the constraints present in the theory
into first and second class\footnote{As a reminder, first/second class constraints are those which do/don't have a weakly vanishing 
Poisson bracket with {\it all} constraints.}.
The algebraic ease is particularly prominent when considering systems with only second class constraints,
as is the case of (multi-)GP electrodynamics.
Last but not least, we note that Dirac's method is a Hamiltonian based one, while Faddeev and Jackiw's is Lagrangian based.
This makes the Faddeev-Jackiw prescription
particularly befitting for dealing with (multi-)GP theories,
which have been formulated and are almost exclusively employed in their Lagrangian formulation.

As with Dirac's original work~\cite{Dirac:1950pj}, Faddeev and Jackiw's proposal~\cite{Faddeev:1988qp} has been extensively
followed upon~\cite{BarcelosNeto:1991wi,Barcelos1,Barcelos,Montani1,Jackiw:1993in,Prescod-Weinstein:2014lua}.
Of particular interest for this work is the path integral formulation of their approach, established in~\cite{Liao, Toms:2015lza}. 
Here, we refer as {\it symplectic quantization} to the quantization procedure derived from the cumulative consideration of~\cite{Faddeev:1988qp,Barcelos1,Barcelos,Liao,Toms:2015lza},
nicely summarized in section 2 of~\cite{Toms:2015lza}.
The outcome of this method is the central object of any quantum field theory: the partition function.

The paper is organized as follows.
We begin with a technical review of multi-GP  in section~\ref{sec:review}.
For clarity, we focus on a particular subset of multi-GP in section~\ref{sec:ourth}
and perform its symplectic quantization in detail in sections~\ref{sec:rewrite}-\ref{sec:output}.
We thus identify two distinct sets of consistency conditions:
\begin{enumerate}
\vspace*{-0.2cm}
\item The already known conditions~\cite{ErrastiDiez:2019trb,ErrastiDiez:2019ttn}, which severely restrict classical, multi-field settings.
\vspace*{-0.2cm}
\item New conditions, which apply in the quantum realm
and affect both single- and multi-field settings.
\end{enumerate}
\vspace*{-0.2cm}
We exemplify the resulting quantization procedure in section~\ref{sec:2simplelimits}.
Section~\ref{sec:newcons} is devoted to the elucidation of the novel quantum consistency conditions.
We conclude with section~\ref{sec:out}, summarizing the results and pointing out possibilities for future work.

\vspace*{0.5cm}
\hspace*{-0.75cm}
\textbf{Conventions.}\\
We work on a $d$-dimensional Minkowski spacetime manifold $\mathcal{M}$, with $d\geq 2$ and the mostly positive metric signature.
Spacetime indices are denoted by the Greek letters $(\mu,\nu,\rho\ldots)$
and raised/lowered with the metric $\eta_{\mu\nu}=\textrm{diag}(-1,1,1,\ldots,1)$ and its inverse $\eta^{\mu\nu}$.
Space indices are denoted by the Latin letters $(i,j,k\ldots)$ and are trivially raised/lowered.
The alphabets $(\alpha,\beta,\ldots)$ label different vector fields.
These vector field labels are trivially raised/lowered.
We employ the standard short-hand notations $\partial_{\mu}f:={\partial f}/{\partial x^{\mu}}$ and $\partial_{i}f:={\partial f}/{\partial x^{i}}$,
where $x^{\mu}$ and $x^{i}$ are spacetime and space local coordinates in $\mathcal{M}$, respectively.
The dot stands for derivation with respect to time: $\dot{f}:=\partial_0f$.
Here, $f$ is any local function $f:\mathcal{M}\rightarrow \mathbb{R}$.
Einstein summation convention applies for all repeated indices and labels throughout the text.

\section{Symplectic quantization}
\label{sec:one}

In this section, we  perform  the detailed symplectic quantization of a family of electrodynamics theories,
all of which describe the dynamics of an arbitrary number $N\in\mathbb{N}$ of GP fields coupled through derivative (self-)interactions.
By definition, the theories here considered  describe multi-field, generalized massive electrodynamics,
whose Lagrangian is manifestly first-order. 

\subsection{Review of multi-GP electrodynamics}
\label{sec:review}

In order to set the notation and contextualize the results obtained in this work, we start with a brief review of our previous work on multi-GP electrodynamics~\cite{ErrastiDiez:2019trb,ErrastiDiez:2019ttn}.
Let $N$ be the number of GP fields $A^{\alpha}=A^\alpha_\mu dx^\mu$, with $\alpha=1,2,\ldots N$.
The most general first-order Lagrangian density,  encoding the dynamics of these GP fields can be written as
\begin{align}
\label{eq:LagTot}
\mathcal{L}_{\textrm{gen}}=\mathcal{L}_{\textrm{kin}}+\mathcal{L}_{\textrm{int}},
\end{align}
where the kinetic piece is canonically normalized
\begin{align}
\label{eq:LagKin}
\mathcal{L}_{\textrm{kin}}=
-\frac{1}{4}A_{\mu\nu}^{\alpha}A^{\mu\nu}_{\alpha}, \qquad
A_{\mu\nu}^{\alpha}:=\partial_{\mu}A_{\nu}^{\alpha}-\partial_{\nu}A_{\mu}^{\alpha},
\end{align}
and the (self-)interaction piece is given by
\begin{align}
\label{eq:LagIntInf}
\mathcal{L}_{\textrm{int}}=\mathcal{L}_{(0)}+\sum_{n=1}^\infty \mathcal{L}_{(n)}.
\end{align}
Here, $\mathcal{L}_{(0)}$ is an arbitrary real smooth function of the GP fields and their field strengths,
\begin{align}
\label{eq:Lag0}
\mathcal{L}_{(0)}=\mathcal{L}_{(0)}(A_{\mu}^{\alpha},A_{\mu\nu}^{\alpha}),
\end{align}
while the factors $\mathcal{L}_{(n)}$  are of the general form
\begin{align}
\label{eq:Lagn}
\mathcal{L}_{(n)}=\mathcal{T}^{\mu_1\ldots \mu_n\nu_1\ldots\nu_n}_{\alpha\ldots\alpha_n}
\partial_{\mu_1}A_{\nu_1}^{\alpha_1}\ldots \partial_{\mu_n}A_{\nu_n}^{\alpha_n},
\end{align}
where the above $\mathcal{T}$ objects are real and smooth and can depend on the GP fields but not on their derivatives:
\begin{align}
\label{eq:Tdepend}
\mathcal{T}^{\mu_1\ldots \mu_n\nu_1\ldots\nu_n}_{\alpha\ldots\alpha_n}=
\mathcal{T}^{\mu_1\ldots \mu_n\nu_1\ldots\nu_n}_{\alpha\ldots\alpha_n}(A_{\mu}^{\alpha}), \qquad
\mathcal{T}^{\mu_1\ldots \mu_n\nu_1\ldots\nu_n}_{\alpha\ldots\alpha_n}\neq
\mathcal{T}^{\mu_1\ldots \mu_n\nu_1\ldots\nu_n}_{\alpha\ldots\alpha_n}(\partial_{\mu}A_{\nu}^{\alpha}).
\end{align}
Therefore, $n$ counts the number of derivative terms of the GP fields present in $\mathcal{L}_{(n\geq 1)}$.
Notice that the Lagrangian $\mathcal{L}_{\textrm{gen}}$ is manifestly first-order.
Namely, it explicitly depends on the GP fields and (powers of) their first derivatives only.
No second- or higher-order derivatives appear.
This feature guarantees that the equations of motion are second-order at most.

In order for the above Lagrangian $\mathcal{L}_{\textrm{gen}}$ to be mathematically well-defined at the classical level,
it must fulfil two necessary and sufficient sets of constraints: (\ref{eq:PrimCons}) and (\ref{eq:SecCons}) below.
The initial GP works~\cite{Tasinato:2014eka,Heisenberg:2014rta} identified (\ref{eq:PrimCons}).
The mathematical procedure was completed in~\cite{ErrastiDiez:2019trb,ErrastiDiez:2019ttn}, with an outcome of (\ref{eq:SecCons}). 
In more detail, (\ref{eq:PrimCons}) enforces the existence of
a second class constraint for every GP field considered.
Such constraints are preserved under time evolution iff (\ref{eq:SecCons}) is fulfilled,
which ensures the existence of another second class constraint per GP field. 
The trivialization of (\ref{eq:SecCons}) for a single GP field implies the automatic existence of the latter second class constraint in this case.
Contrastively, multi-field (and therefore non-abelian) settings are severely restricted by (\ref{eq:SecCons}).

The first set of constraints has been referred to as {\it primary constraint enforcing relations} and is given by
\begin{align}
\label{eq:PrimCons}
\frac{\partial^2 \mathcal{L}_{\textrm{gen}}}{\partial \dot{A}_{0}^{\alpha}\partial \dot{A}_{\mu}^{\beta}}\overset{!}{=}0.
\end{align}
This has two drastic consequences on $\mathcal{L}_{\textrm{gen}}$.
On the one hand, it truncates the sum over $n$ in (\ref{eq:LagIntInf}) at $n=d$, so that the interaction piece reduces to
\begin{align}
\label{eq:LagInt4}
\mathcal{L}_{\textrm{int}}=\mathcal{L}_{(0)}+\sum_{n=1}^d \mathcal{L}_{(n)}.
\end{align}
On the other hand, it forces a certain form on the $\mathcal{T}$ objects in $\mathcal{L}_{(n\geq 2)}$, albeit without fully fixing them.
The interested reader can consult the form of such $\mathcal{T}$'s, for the particular case when $d=4$, in equations (21)-(23) of~\cite{ErrastiDiez:2019trb}.

The second set of constraints, the so-called  {\it secondary constraint enforcing relations}, is
\begin{align}
\label{eq:SecCons}
\frac{\partial^2 \mathcal{L}_{\textrm{gen}}}{\partial\dot{A}_0^{\alpha}\partial A_0^{\beta}}
-\frac{\partial^2 \mathcal{L}_{\textrm{gen}}}{\partial\dot{A}_0^{\beta}\partial A_0^{\alpha}}\overset{!}{=}0.
\end{align}
The above further restricts the form of the $\mathcal{T}$ objects in all $\mathcal{L}_{(n\geq1)}$,
although it still does not completely determine them.
Owing to the very significant complexity of both $\mathcal{L}_{\textrm{gen}}$ and (\ref{eq:SecCons}),
the latter has not yet been exhaustively implemented, even in the particular $d=4$ case.
Namely, to date there is no complete list of $\mathcal{T}$'s that simultaneously satisfy (\ref{eq:PrimCons}) and (\ref{eq:SecCons}).
Therefore, for the time being, (\ref{eq:SecCons}) is to be viewed as an essential {\it classical consistency condition}
that must be fulfilled in any multi-GP electrodynamics one may wish to consider.
Particular examples of $\mathcal{T}$'s  in $\mathcal{L}_{(1)}$ and $\mathcal{L}_{(2)}$ 
that satisfy both (\ref{eq:PrimCons}) and (\ref{eq:SecCons}) have been proposed in $d=4$~\cite{ErrastiDiez:2019trb,ErrastiDiez:2019ttn}.

\subsection{The targeted multi-GP electrodynamics}
\label{sec:ourth}

In the present work, we will restrict computational attention to the following subset of interactions
within the above described $\mathcal{L}_{\textrm{gen}}$ massive electrodynamics theory:
\begin{align}
\label{eq:Lagchosen}
\mathcal{L}_{(0)}=-\frac{1}{2}m^2A_{\mu}^{\alpha}A^{\mu}_{\alpha}+
cA_{\mu}^{\alpha}A^{\mu}_{\beta}A_{\nu}^{\beta}A^{\nu}_{\alpha}, \qquad
\mathcal{L}_{(n\geq2)}=0,
\end{align}
with $m\in \mathbb{R}^+$ the (hard) mass of the GP fields (chosen to be the same for all GP fields for simplicity)
and $c\in\mathbb{R}$ a dimensionless constant.
We will consider all the terms in $\mathcal{L}_{(1)}$.
Specifying a certain $\mathcal{L}_{(0)}$ is necessary in order to explicitly (as opposed to formally) carry out the symplectic quantization procedure.
We here choose to consider the standard mass term for the GP fields, originally proposed in~\cite{Proca1,Proca2},
as well as the quartic interactions among the GP fields.
The latter are the simplest (self-)interactions for the massive vector fields, yet they are interesting in their own right.
Remarkably, they have been shown to admit a non-Wilsonian ultraviolet completion~\cite{Dvali:2010jz}.
They lead to time-dependent solitonic solutions~\cite{Nobre}.
Recently, such terms have attracted attention in the context of Proca Stars as well~\cite{Minamitsuji:2018kof,Herdeiro:2021lwl}.
Here, we introduced the constant $c$ to straightforwardly keep track of subsequent contributions stemming from these quartic interactions.
We will comment on the non-trivialities involved in the extensions of (\ref{eq:Lagchosen}) with $\mathcal{L}_{(n\geq 2)}\neq 0$ shortly.
At last, we will include external sources $J^\mu_\alpha$.
All in all, we shall consider the particular multi-GP electrodynamics theories encoded in
\begin{align}
\label{eq:ourLag}
\mathcal{L}_{\textrm{par}}=-\frac{1}{4}A_{\mu\nu}^{\alpha}A^{\mu\nu}_{\alpha}-
\frac{1}{2}m^2A_{\mu}^{\alpha}A^{\mu}_{\alpha}+
cA_{\mu}^{\alpha}A^{\mu}_{\beta}A_{\nu}^{\beta}A^{\nu}_{\alpha}+
\mathcal{T}^{\mu\nu}_{\alpha}\partial_{\mu}A_{\nu}^{\alpha}+A_\mu^\alpha J^\mu_\alpha,
\end{align}
where the objects $\mathcal{T}^{\mu\nu}_{\alpha}$ are required to satisfy the classical consistency condition (\ref{eq:SecCons}),
with $\mathcal{L}_{\textrm{gen}}$ replaced by $\mathcal{L}_{\textrm{par}}$.

Here, the $A_{\mu}^{\alpha}$'s are the {\it generalized coordinates}
(that is, the a priori independent degrees of freedom in terms of which the electrodynamics theories of our interest are described):
\begin{align}
\label{eq:naiveGC}
Q=\{A_{\mu}^{\alpha}\}.
\end{align}
The generalized coordinates span the {\it configuration space} of the theories, which in this case is $dN$-dimensional.
The time derivatives of the generalized coordinates are the {\it generalized velocities}:
\begin{align}
\label{eq:naiveGV}
\dot{Q}=\{\dot{A}_{\mu}^{\alpha}\}.
\end{align}

Upon a space-time decomposition, (\ref{eq:ourLag}) becomes
\begin{align}
\label{eq:ourLagSandT}
\begin{array}{lllllllll}
\displaystyle
\mathcal{L}_{\textrm{par}}=&\frac{1}{2}\dot{A}_{i}^{\alpha}\dot{A}^{i}_{\alpha}+
\dot{A}_{i}^{\alpha}\partial^{i}A_{\alpha}^0-
\frac{1}{2}(\partial^{i}A_{\alpha}^0)\partial_{i}A_0^{\alpha}-
\frac{1}{4}A_{ij}^{\alpha}A^{ij}_{\alpha} \vspace*{0.2cm}\\
&-\frac{1}{2}m^2\left(A_0^{\alpha}A^0_{\alpha}+A_i^{\alpha}A^i_{\alpha}\right)+
c\left(A_0^{\alpha}A^0_{\beta}A^{\beta}_0A^0_{\alpha}+
2A_0^{\alpha}A^0_{\beta}A^{\beta}_{i}A^{i}_{\alpha}+
A_{i}^{\alpha}A^{i}_{\beta}A^{\beta}_{j}A^{j}_{\alpha}\right) \vspace*{0.2cm}\\
&+\mathcal{T}^{00}_{\alpha}\dot{A}_0^{\alpha}+\mathcal{T}^{0i}_{\alpha}\dot{A}^{\alpha}_{i}
+\mathcal{T}^{i0}_{\alpha}\partial_{i}A_0^{\alpha}+\mathcal{T}^{ij}_{\alpha}\partial_{i}A_{j}^{\alpha}+A_0^\alpha J^0_\alpha +A_i^\alpha J^i_\alpha,
\end{array}
\end{align}
where, for the convenience of the reader, we have placed the terms coming from $\mathcal{L}_{\textrm{kin}}$, $\mathcal{L}_{(0)}$ and $\mathcal{L}_{(1)}$
(plus the coupling to the external sources) in the first, second and third lines, respectively.
The classical consistency condition for the above explicitly reads
\begin{align}
\label{eq:ourclasscond}
\overline{\partial}^0_\beta\mathcal{T}^{00}_\alpha -\overline{\partial}^0_\alpha\mathcal{T}^{00}_\beta \overset{!}{=}0,
\end{align}
where we have introduced the short-hand
\begin{align}
\label{eq:barder}
\overline{\partial}^\mu_\alpha:=\frac{\partial}{\partial A_\mu^\alpha}.
\end{align}

\subsection{Input for the iterative procedure}
\label{sec:rewrite}

The symplectic quantization method can only be employed on Lagrangian densities which are linear in the generalized velocities.
Namely, Lagrangian densities of the form
\begin{align}
\label{eq:FJLagform}
\mathcal{L}=\theta \cdot \dot{Q} +\widehat{\mathcal{L}}, 
\end{align}
where $\theta$ and $\widehat{\mathcal{L}}$ are functions of the generalized coordinates $Q$ but not of the generalized velocities $\dot{Q}$.
$\theta$ is known as the {\it canonical one-form}.
Upon termination of the symplectic quantization iterative procedure, $\widehat{\mathcal{L}}$ is minus the Hamiltonian density.

Clearly, (\ref{eq:ourLagSandT}) is not of the above form.
Indeed, $\mathcal{L}_{\textrm{par}}$ contains quadratic terms in the generalized velocities.
These stem from $\mathcal{L}_{\textrm{kin}}$.
In order to bring (\ref{eq:ourLagSandT}) to the desired form (\ref{eq:FJLagform}),
we will extend the configuration space of our theory,
by declaring the {\it canonical momenta} $p^{\mu}_{\alpha}$ (with respect to $A_\mu^\alpha$) generalized coordinates as well:
\begin{align}
\label{eq:notsonariveGC}
Q=\{A_{\mu}^{\alpha},p^\mu_\alpha\}.
\end{align}
At this point, we thus consider a configuration space that is $2dN$-dimensional,
with the canonical momenta given by
\begin{align}
\label{eq:Ps}
p^0_\alpha:=\frac{\partial \mathcal{L}_{\textrm{par}}}{\partial \dot{A}_0^\alpha}=\mathcal{T}^{00}_\alpha, \qquad
p^i_\alpha:=\frac{\partial \mathcal{L}_{\textrm{par}}}{\partial \dot{A}_i^\alpha}=\dot{A}^i_\alpha+\partial^i A^0_\alpha+\mathcal{T}^{0i}_\alpha.
\end{align}

It is of utmost importance to make the following two observations. First, the canonical momenta $p^i_\alpha$ depend on (some of) the generalized velocities $\dot{Q}$, while the canonical momenta $p^0_\alpha$ do not.
The fact that $p^0_\alpha\neq p^0_\alpha(\dot{Q})$ is a direct consequence of the primary constraint enforcing relations (\ref{eq:PrimCons})
and it implies that we must view
\begin{align}
\label{eq:P0cons}
\varphi_\alpha:=p^0_\alpha-\mathcal{T}^{00}_\alpha\overset{!}{=}0
\end{align}
as a set of $N$ number of (functionally independent) constraints that must be appropriately accounted for in our considered theories.
This can be readily done via Lagrange multipliers $\lambda^\alpha$,
which we must regard as further generalized coordinates:
\begin{align}
\label{eq:notnaiveGC}
Q=\{A_{\mu}^{\alpha},p^\mu_\alpha,\lambda^\alpha\}.
\end{align}
We thus settle for a $(2d+1)N$-dimensional configuration space associated to (\ref{eq:ourLagSandT})
with views to performing the symplectic quantization of the theories.

Second, we notice that the second set of equalities in (\ref{eq:Ps}) forms a system of $(d-1)N$ number of linearly independent equations.
Such linear independence is guaranteed by construction~\cite{ErrastiDiez:2019trb} for all electrodynamics theories reviewed in the previous section \ref{sec:review}.
Further, in the particular case at hand, it is straightforward to solve this system for $\dot{A}^i_\alpha$ in terms of $(p^i_\alpha, A_\mu^\alpha)$:
\begin{align}
\label{eq:velseqmomen}
\dot{A}^i_\alpha=p^i_\alpha-\partial^i A^0_\alpha-\mathcal{T}^{0i}_\alpha.
\end{align}

The situation becomes more involved if $\mathcal{L}_{(n\geq 2)}\neq 0$.
When $\mathcal{L}_{(2)}\neq 0$ with $\mathcal{L}_{(n\geq 3)}= 0$, the aforementioned linear independence ensures
a unique solution $\dot{A}^i_\alpha=\dot{A}^i_\alpha(p^i_\alpha, A_\mu^\alpha)$ exists.
Then, the difficulty amounts to the algebraic effort required for its explicit determination.
Whenever $\mathcal{L}_{(n\geq 3)}\neq 0$, we encounter a polynomial in $\dot{A}^i_\alpha$ of order $(n-1)$ on the right-hand side of the second set of equalities in (\ref{eq:Ps}).
We are thus confronted with a setting where the inversion of the generalized velocities in terms of the canonical momenta (and the generalized coordinates) is multivalued. 
This looks like a worse problem than it actually is: the complication is a technical --- as opposed to a fundamental --- one
and was elegantly resolved in~\cite{Avraham:2014twa} by defining a generalized notion for the Legendre transform.
The increased algebraic effort associated with choosing $\mathcal{L}_{(n\geq 2)}\neq 0$ is notorious, but certainly not insurmountable,
and would obscure the transcendence of our results.
For this reason, we have opted to set $\mathcal{L}_{(n\geq 2)}= 0$ in this work.

Overall, the reconsideration of (\ref{eq:ourLagSandT}) such that (\ref{eq:notnaiveGC}) are the generalized coordinates yields,
upon minor algebraic effort employing (\ref{eq:velseqmomen}), a Lagrangian density of the desired form (\ref{eq:FJLagform}),
with $\theta=\{p^\mu_\alpha,0,\varphi_\alpha\}$ and
\begin{align}
\label{eq:ourLagprime}
\begin{array}{lllllll}
\displaystyle
\widehat{\mathcal{L}}=&-\frac{1}{2}p^i_\alpha p^\alpha_i-p^i_\alpha\partial_iA_0^\alpha-\frac{1}{4}A_{ij}A^{ij} \vspace*{0.2cm}\\
&-\frac{1}{2}m^2\left(A_0^{\alpha}A^0_{\alpha}+A_i^{\alpha}A^i_{\alpha}\right)+
c\left(A_0^{\alpha}A^0_{\beta}A^{\beta}_0A^0_{\alpha}+
2A_0^{\alpha}A^0_{\beta}A^{\beta}_{i}A^{i}_{\alpha}+
A_{i}^{\alpha}A^{i}_{\beta}A^{\beta}_{j}A^{j}_{\alpha}\right) \vspace*{0.2cm}\\
&+(p^\alpha_i+\partial_iA_0^\alpha)\mathcal{T}^{0i}_\alpha+\frac{1}{2}\mathcal{T}^{0i}_\alpha\mathcal{T}_{0i}^\alpha
+\mathcal{T}^{i0}_{\alpha}\partial_{i}A_0^{\alpha}+\mathcal{T}^{ij}_{\alpha}\partial_{i}A_{j}^{\alpha} +A_0^\alpha J^0_\alpha + A_i^\alpha J^i_\alpha, 
\end{array}
\end{align}
where, once more for the convenience of the reader,
we have placed the terms coming from $\mathcal{L}_{\textrm{kin}}$, $\mathcal{L}_{(0)}$ and $\mathcal{L}_{(1)}$
(plus the coupling to the external sources) in the first, second and third lines, respectively.
Of course, the classical consistency conditions (\ref{eq:ourclasscond}) must be fulfilled in this rewriting as well.

Here, it is important to note note that we have viewed the essential terms enforcing the constraints (\ref{eq:P0cons})
via Lagrange multipliers as belonging within the symplectic part of the Lagrangian, i.e.~the first term in (\ref{eq:FJLagform}).
This is because the Lagrange multipliers are arbitrary, so we can enforce (\ref{eq:P0cons}) via their time derivatives just as well.
In other words, we can incorporate the constraints (\ref{eq:P0cons}) to our electrodynamics theories in two physically equivalent ways:
adding either ${\lambda}^\alpha\varphi_\alpha$ or $\dot{\lambda}^\alpha\varphi_\alpha$ to (\ref{eq:ourLagSandT}).
The first way is followed in Dirac-based standard quantization procedures, whereas the second way is a cornerstone to the symplectic quantization methodology.
The interested reader can consult~\cite{Barcelos} for a detailed exploration of the said two manners to incorporate constraints,
as well as a proof of their physical equivalence.
In the present work, we have of course elected the second option.

An important technical remark is as follows.
The expert reader may here worry that we are overlooking the prescription in~\cite{Seiler} for field theories.
Namely, that we may be missing out on unveiling purely spatial consistency conditions, since these can only be found by introducing
$d$ number of Lagrange multipliers per constraint, in the form $\partial_\mu \lambda^{\mu\alpha}\varphi_\alpha$.
We have explicitly checked that no such spatial conditions apply to our considered settings (\ref{eq:ourLag}) and, a posteriori,
have opted for alleviating the algebraic presentation throughout the text
by only introducing one Lagrange multiplier per constraint: $\dot{\lambda}^\alpha\varphi_\alpha$.
The inclusion of all $d$ Lagrange multipliers leads to the generation of  functionally dependent $(d-1)$ number of constraints at the first iteration,
given by $\partial_i \varphi_\alpha\overset{!}{=}0$, which are simply redundant.

For completeness, we point out that our above manipulation of (\ref{eq:ourLag}), or equivalently of (\ref{eq:ourLagSandT}),
to bring it into the FJ form (\ref{eq:FJLagform}) is not the only possible one.
It is the one employed in~\cite{Toms:2015lza} for the symplectic quantization of Proca electrodynamics
and therefore our forthcoming results are most easily compared to this reference, in the appropriate limit.
It is worth noting that~\cite{Prescod-Weinstein:2014lua} also promotes canonical momenta and Lagrange multipliers to additional generalized coordinates
for the quantization of Proca electrodynamics. 
However, this work is primarily concerned with the introduction of a distinct, albeit Faddeev-Jackiw-based, quantization procedure.
Therefore, a step-wise comparison of our work to~\cite{Prescod-Weinstein:2014lua} is not possible.
There is another possibility, which was exploited in~\cite{Pimentel:2015mza},
also in the context of the symplectic quantization of Proca electrodynamics.
In this reference, the theory is first manipulated to enjoy a $U(1)$ gauge symmetry.
This is achieved through the suitable inclusion of an additional scalar field,
in a procedure that in some contexts is referred to as the {\it St\"uckelberg mechanism}, originally proposed in~\cite{Stueckelberg:1900zz}.
(We refer the interested reader to~\cite{Ruegg:2003ps} for a compelling modern review of this mechanism.)
Afterwards, the Proca and scalar fields, together with their canonical momenta are regarded as the generalized coordinates and the symplectic quantization method is employed.
While it is possible to proceed in an analogous manner for our considered electrodynamics theories (\ref{eq:ourLag}), this is algebraically more cumbersome.
With simplicity in mind, we have opted for quantizing the theories as they are, with no gauge symmetry at all.
We stress that the said two distinct manners in which a Lagrangian can be brought into the form (\ref{eq:FJLagform})
are explicitly shown to yield the same physics in~\cite{Nogueira:2018jdm} for the non-trivial case of Podolsky electrodynamics~\cite{Podolsky:1942zz,Podolsky:1944zz}.
For clarity, we point out that the authors of~\cite{Nogueira:2018jdm} refer to the aforementioned enlargement of the configuration space and to the St\"uckelberg mechanism as
{\it reduced order formalism} and {\it Ostrogradsky prescription}, respectively.
The latter name alludes to the original paper~\cite{Ostrogradsky:1850fid}, but employs the modern understanding developed in~\cite{TSChang,Pais:1950za}.

\subsection{First iteration}
\label{sec:it1}

The first step in the symplectic quantization prescription amounts to the calculation of the so-called {\it symplectic two-form} $\Omega$,
a totally anti-symmetric square matrix, whose components are given by
\begin{align}
\label{eq:sym2formdef}
\Omega_{mn}:=\frac{\delta\theta^\prime_n }{\delta Q^m}-\frac{\delta\theta_m}{\delta {Q^\prime}^{\,n}},
\end{align}
where $m,n=1,2,\ldots, (2d+1)N$ label the individual elements in $\theta=\{p^\mu_\alpha,0,\varphi_\alpha\}$ and $Q$ in (\ref{eq:notnaiveGC}).
The symplectic two-form is defined on a constant time hypersurface $\Sigma \subset \mathcal{M}$.
The non-primed quantities $(\theta, Q)$ are to be understood as evaluated at some point $x=(t^\ast, x^i) \in \Sigma$, with $t^\ast$ an arbitrary but fixed time;
while their primed counterparts $(\theta^\prime, Q^\prime)$ are to be understood as evaluated at some other point $x^\prime =(t^\ast,{x^\prime}^{\,i}) \in\Sigma$.
We can succinctly spell out $\Omega$ as
\begin{align}
\label{eq:ouromega1}
\displaystyle
\Omega=\left(
\begin{array}{ccccccc}
0 &\quad -\delta^\mu_\nu \delta^\beta_\alpha &\quad -\overline{\partial}^\mu_\alpha \mathcal{T}^{00}_\beta \vspace*{0.2cm}\\
\delta^\nu_\mu\delta^\alpha_\beta &\quad 0 &\quad \delta^0_\mu\delta^\alpha_\beta \vspace*{0.2cm}\\
\overline{\partial}^\nu_\beta\mathcal{T}^{00}_\alpha &\quad -\delta^0_\nu\delta^\beta_\alpha &\quad 0
\end{array}
\right)\delta^{d-1}(x^i-{x^\prime}^{\, i}).
\end{align}

Next, we need to determine whether the above symplectic two-form is singular or not.
The calculation of the determinant is subtle, so we will carry it out explicitly.
To this aim, we will make use of Schur's identity.
Namely, given any square matrix $M$ that admits a block decomposition of the form
\begin{align}
\label{eq:Mdecom}
\displaystyle
M=\left(
\begin{array}{ccccc}
M_1 &\quad M_2 \vspace*{0.2cm}\\
M_3 &\quad M_4
\end{array}
\right), 
\end{align}
such that $M_1$ and $M_4$ are square and $M_1$ is invertible, its determinant can be computed as
\begin{align}
\label{eq:detM}
\textrm{det}(M)=\textrm{det}(M_1)\textrm{det}(M_4-M_3M_1^{-1}M_2).
\end{align}
Notice that Schur's identity does {\it not} require $M_2$ and $M_3$ to be square.
Upon the identifications $M=\Omega$,
\begin{align}
\label{eq:idents}
\displaystyle
\begin{array}{llllllll}
&M_1=\left(
\begin{array}{ccccc}
0 &\quad -\delta^\mu_\nu \delta^\beta_\alpha \vspace*{0.2cm}\\
\delta^\nu_\mu\delta^\alpha_\beta &\quad 0
\end{array}
\right)\delta^{d-1}(x^i-{x^\prime}^{\, i}), 
&\qquad M_2=\left(
\begin{array}{cccc}
-\overline{\partial}^\mu_\alpha \mathcal{T}^{00}_\beta \vspace*{0.2cm}\\
\delta^0_\mu\delta^\alpha_\beta
\end{array}
\right)\delta^{d-1}(x^i-{x^\prime}^{\, i}), \vspace*{0.4cm}\\
&M_3=\left(
\begin{array}{cccc}
\overline{\partial}^\nu_\beta\mathcal{T}^{00}_\alpha &\quad -\delta^0_\nu\delta^\beta_\alpha
\end{array}
\right)\delta^{d-1}(x^i-{x^\prime}^{\, i}),
&\qquad M_4=0
\end{array}
\end{align}
and noting that
\begin{align}
\label{eq:M1props}
\textrm{det}(M_1)=1, \qquad M_1^{-1}=M_1,
\end{align}
we easily arrive at
\begin{align}
\label{eq:detOmega1}
\textrm{det}(\Omega)=\textrm{det}(-M_3M_1M_2)=
\textrm{det}\left[\left(\overline{\partial}^0_\alpha\mathcal{T}^{00}_\beta-\overline{\partial}^0_\beta\mathcal{T}^{00}_\alpha\right)\delta^{d-1}(x^i-{x^\prime}^{\, i})\right].
\end{align}
By virtue of the classical consistency conditions (\ref{eq:ourclasscond}), the above determinant vanishes.
The symplectic two-form $\Omega$ in (\ref{eq:ouromega1}) is therefore singular.
Its singularity implies the existence of further constraints, beyond the already unveiled ones in (\ref{eq:P0cons}).
Before calculating these additional constraints, we reflect upon (\ref{eq:detOmega1}).

For just a moment, suppose that we would not have been aware of the classical consistency conditions (\ref{eq:ourclasscond}) from the very beginning.
In such a case, at this point we would have {\it derived} (\ref{eq:ourclasscond}) from (\ref{eq:detOmega1}).
This is because the singularity of the symplectic two-form is indispensable for the correct postulation of any electrodynamics theory
and thus for our considered particular theory (\ref{eq:ourLag}) too.
For instance, it is well known that Proca electrodynamics is associated with two (second-class) constraints.
The first such constraint amounts to the independence of the action from $p^0$ or, equivalently, from $\dot{A}_0$.
The second constraint exists iff\footnote{This statement will become clear shortly, in (\ref{eq:moreconsdef}).} the symplectic two-form vanishes.
In the Proca case, this vanishing is automatic.
We now turn to the more general GP case.
Since all (multi-)GP are non-linear extensions of Proca electrodynamics, they must have its same constraint algebraic structure:
each GP field must be associated with two (second-class) constraints.
The first set is that in (\ref{eq:P0cons}).
The second set exists iff (\ref{eq:detOmega1}) is zero,
which uniquely and straightforwardly implies (\ref{eq:ourclasscond}).
Therefore, at this point we have obtained the following important side-result: 
an independent derivation of the classical consistency conditions applying to all multi-GP electrodynamics theories,
which were originally disclosed in~\cite{ErrastiDiez:2019trb,ErrastiDiez:2019ttn} following a different approach, \`a la Dirac.

As a first step in the determination of the necessary additional constraints in our considered generalized massive electrodynamics theories,
we compute the zero modes of $\Omega$ in (\ref{eq:ouromega1}).
The number of linearly independent zero modes that $\Omega$ admits is equal to
\begin{align}
\label{eq:numb0modes}
\textrm{dim}(\Omega)-\textrm{rank}(\Omega)=(2d+1)N-2dN=N.
\end{align}
The above rank readily follows from the observation that (\ref{eq:detOmega1}) identically vanishes for a single GP field, together with (\ref{eq:M1props}).
The $N$ linearly independent zero modes of $\Omega$ are of the generic form $\gamma_\alpha=(u^\alpha_\mu,v^\mu_\alpha,w^\alpha)$
and fulfill that their left multiplication with $\Omega$ vanishes\footnote{This is but a harmless choice.
It is also possible to choose to define the zero modes as those column vector, whose right multiplication with $\Omega$ yields zero.
Here, we opt for the left multiplication convention also employed in~\cite{Toms:2015lza}, with the constant goal to make our results easily comparable to
the limiting scenario of Proca electrodynamics worked out in this reference.}.
This vanishing implies
\begin{align}
\label{eq:0modeform}
u_0^\alpha=-w^\alpha, \qquad u_i^\alpha=0, \qquad v^0_\alpha=-w^\beta\overline{\partial}^0_\beta\mathcal{T}^{00}_\alpha, \qquad
v^i_\alpha=-w^\beta\overline{\partial}^i_\alpha \mathcal{T}^{00}_\beta
\end{align}
and we have the freedom to choose the $w^\alpha$ components.
A simple consistent choice amounts to setting
\begin{align}
w^\alpha=(0,0,\ldots,0,-1,0,0,\ldots,0)=:-\mathbb{I}^\alpha,
\end{align}
where the non-zero entry is in the $\alpha$-th position.
All in all, we shall consider the following zero modes of $\Omega$:
\begin{align}
\label{eq:gammas}
\gamma_\alpha =\left(\delta^0_\mu\mathbb{I}^\alpha,\mathbb{I}^\beta(\delta^\mu_0\overline{\partial}^0_\beta\mathcal{T}^{00}_\alpha+
\delta^\mu_i\overline{\partial}^i_\alpha\mathcal{T}^{00}_\beta),-\mathbb{I}^\alpha\right).
\end{align}

There are as many new constraints as linearly independent zero modes.
These additional constraints $\widetilde{\varphi}_\alpha$ can be determined employing the above zero modes according to the formula
\begin{align}
\label{eq:moreconsdef}
\widetilde{\varphi} := \gamma \cdot \frac{\delta \widehat{\mathcal{L}}}{\delta Q}\overset{!}{=}0,
\end{align}
with the generalized coordinates $Q$, the non-symplectic part of the Lagrangian density $\widehat{\mathcal{L}}$ and the zero modes $\gamma_\alpha$
as given in (\ref{eq:notnaiveGC}), (\ref{eq:ourLagprime}) and (\ref{eq:gammas}), respectively.
It is easy to verify that the above constraints are explicitly given by
\begin{align}
\label{eq:newconsexpl}
\begin{array}{lllllll}
\widetilde{\varphi}_\alpha=&-m^2A^0_\alpha+2cA^0_\beta\left(A^\beta_0A^0_\alpha+2A^\beta_iA^i_\alpha\right)
+\left(p_i^\beta+2\partial_iA_0^\beta+\mathcal{T}^\beta_{0i}\right)
\overline{\partial}^0_\alpha \mathcal{T}^{0i}_\beta
+\left(\partial_iA_j^\beta\right)\overline{\partial}^0_\alpha \mathcal{T}^{ij}_\beta
\vspace*{0.2cm}\\
&+\left(p_i^\beta+\partial_iA_0^\beta+\mathcal{T}^\beta_{0i}\right)
\overline{\partial}^i_\beta \mathcal{T}^{00}_\alpha
+\partial_i p^i_\alpha
-\partial_i\left(\mathcal{T}^{0i}_\alpha+\mathcal{T}^{i0}_\alpha\right)
+J^0_\alpha\overset{!}{=}0.
\end{array}
\end{align}

Henceforth, it is essential to only consider the functionally independent constraints.
As was the case for (\ref{eq:P0cons}) earlier on, the functional independence of the above constraints is also ensured by construction~\cite{ErrastiDiez:2019trb}.
Therefore, all $N$ number of constraints in (\ref{eq:newconsexpl}) must be taken into account.
We redirect the interested reader to section IID in~\cite{Diaz:2017tmy} for an astute methodology to deal with (almost all) scenarios
where there is no functional independence among the constraints.
It is worth noting that this reference contains enlightening examples as well.

The (functionally independent) constraints (\ref{eq:newconsexpl}) are to be incorporated through new Lagrange multipliers $\widetilde{\lambda}^\alpha$.
The novel Lagrange multipliers must be viewed as further generalized coordinates, so that the configuration space of our electrodynamics theories is now spanned by
\begin{align}
\label{eq:GCfinal}
Q=\{A_{\mu}^{\alpha},p^\mu_\alpha,\lambda^\alpha,\widetilde{\lambda}^\alpha\}
\end{align}
and is $2(d+1)N$-dimensional.
Following our remarks below (\ref{eq:ourLagprime}), we include the terms $\widetilde{\varphi}_\alpha\dot{\widetilde{\lambda}}^\alpha$ to our Lagrangian density.
This is of the required form (\ref{eq:FJLagform}), with
\begin{align}
\label{eq:thetafinal}
\theta=\{p^\mu_\alpha,0,\varphi_\alpha,\widetilde{\varphi}_\alpha\}
\end{align}
and $\widehat{\mathcal{L}}$ as in (\ref{eq:ourLagprime}).
Once again, there is no need to include $d$ number of Lagrange multipliers per constraint $\partial_\mu \widetilde{\lambda}^{\mu\alpha}\widetilde{\varphi}_\alpha$,
as generally required for field theories~\cite{Seiler}. 
This is because no constraints arise at the second iteration, as we shall immediately show.

\subsection{Second iteration}
\label{sec:it2}

We proceed to calculate the symplectic two-form $\Omega$ associated to the above obtained Lagrangian density.
We do so according to the definition in (\ref{eq:sym2formdef}), but this time with $n,m=1,2,\ldots,2(d+1)N$
referring to the components of $Q$ and $\theta$ in (\ref{eq:GCfinal}) and (\ref{eq:thetafinal}), respectively.
The result is
\begin{align}
\label{eq:omegafinal}
\displaystyle
\Omega=\left(
\begin{array}{ccccccc}
0 &\quad -\delta^\mu_\nu \delta^\beta_\alpha &\quad -\overline{\partial}^\mu_\alpha \mathcal{T}^{00}_\beta &\quad X^\mu_{\alpha\beta}\vspace*{0.2cm}\\
\delta^\nu_\mu\delta^\alpha_\beta &\quad 0 &\quad \delta^0_\mu\delta^\alpha_\beta &\quad -Y^\alpha_{\mu\beta} \vspace*{0.2cm}\\
\overline{\partial}^\nu_\beta\mathcal{T}^{00}_\alpha &\quad -\delta^0_\nu\delta^\beta_\alpha &\quad 0 &\quad 0 \vspace*{0.2cm}\\
-X^{\prime\nu}_{\beta\alpha} &\quad Y^{\prime\beta}_{\nu\alpha} &\quad 0 &\quad 0 
\end{array}
\right)\delta^{d-1}(x^i-{x^\prime}^{\, i}),
\end{align}
where we have introduced
\begin{align}
\label{eq:XYdef}
\displaystyle
\begin{array}{rlllll}
X^{0}_{\alpha\beta}&\hspace*{-0.3cm}:=&\hspace*{-0.3cm}
m^2\delta_{\alpha}^{\beta}-2c\left(A_0^\gamma A^0_\gamma \delta_{\alpha}^{\beta}
+2A_\mu^\beta A^\mu_\beta\right)
+\left(\overline{\partial}^0_\beta\mathcal{T}^{0i}_\gamma
+\overline{\partial}^i_\gamma\mathcal{T}^{00}_\beta\right)\overline{\partial}^0_\alpha\mathcal{T}^\gamma_{0i}
+\Big(p_i^\gamma+2\partial_i^\prime A_0^\gamma+\mathcal{T}^\gamma_{0i}\Big)
\overline{\partial}^0_\alpha\overline{\partial}^0_\beta\mathcal{T}^{0i}_\gamma
\vspace*{0.2cm}\\
&&\hspace*{-0.3cm}
+\Big(\partial_i^\prime A_j^\gamma\Big)
\overline{\partial}^0_\alpha\overline{\partial}^0_\beta\mathcal{T}^{ij}_\gamma
+\Big(p_i^\gamma+\partial_i^\prime A_0^\gamma+\mathcal{T}^\gamma_{0i}\Big)
\overline{\partial}^0_\alpha \overline{\partial}^i_\gamma \mathcal{T}^{00}_\beta
-\partial_i^\prime\overline{\partial}^0_\alpha
\left(\mathcal{T}^{0i}_\beta+\mathcal{T}^{i0}_\beta\right)
-2\partial_i\overline{\partial}^0_\beta\mathcal{T}^{0i}_\alpha
-\partial_i\overline{\partial}^i_\alpha\mathcal{T}^{00}_\beta,
\vspace*{0.4cm}\\
X^i_{\alpha\beta}&\hspace*{-0.3cm}:=&\hspace*{-0.3cm}
4c\left(A^0_\alpha A^i_\beta -A_0^\gamma A^i_\gamma\delta_{\alpha}^{\beta}\right)
+\left(\overline{\partial}^0_\beta\mathcal{T}_\gamma^{0j}+\overline{\partial}^j_\gamma\mathcal{T}_\beta^{00}\right)\overline{\partial}^i_\alpha\mathcal{T}^\gamma_{0j}
+\left(p_j^\gamma+2\partial_j^\prime A_0^\gamma +\mathcal{T}^\gamma_{0j}\right)\overline{\partial}^0_\beta\overline{\partial}^i_\alpha\mathcal{T}_\gamma^{0j}
\vspace*{0.2cm}\\
&&\hspace*{-0.3cm}
+\Big(\partial^\prime _j A_k^\gamma \Big)\overline{\partial}^0_\beta\overline{\partial}^i_\alpha\mathcal{T}_\gamma^{jk}
+\left(p_j^\gamma+\partial_j^\prime A_0^\gamma +\mathcal{T}^\gamma_{0j}\right)\overline{\partial}^i_\alpha\overline{\partial}^j_\gamma\mathcal{T}_\beta^{00}
-\partial_j\overline{\partial}^0_\beta\mathcal{T}^{ji}_\alpha
-\partial_j^\prime\overline{\partial}^i_\alpha\left(\mathcal{T}^{0j}_\beta+\mathcal{T}^{j0}_\beta\right),  \vspace*{0.4cm}\\
Y^{\alpha}_{\mu\beta}&\hspace*{-0.3cm}:=&\hspace*{-0.3cm}
\delta^i_\mu\left(\delta_\alpha^\beta \partial_i+\overline{\partial}^\alpha_i\mathcal{T}^{00}_\beta+\overline{\partial}^0_\beta\mathcal{T}^\alpha_{0i} \right).
\end{array}
\end{align}
The primed counterparts $X^{\prime\mu}_{\alpha\beta}$ and $Y^{\prime\alpha}_{\mu\beta}$ follow from
replacing $\partial_i\leftrightarrow (-)\partial_i^\prime$ everywhere in the above expressions,
with $\partial_i^\prime$ the short-hand for derivation with respect to ${x^\prime}^{\, i}$.
The minus sign applies only for those partial derivatives $\partial_i^{(\prime)}$ that act on the Dirac delta $\delta^{d-1}(x^i-{x^\prime}^{\, i})$.
Namely, the first term on the right-hand side of $Y^{\alpha}_{\mu\beta}$.
We note the additional components in (\ref{eq:omegafinal}), as compared to (\ref{eq:ouromega1}) before.
These stem directly from the newly found constraints in (\ref{eq:newconsexpl}).
We stress that, generically, $X^\mu_{\alpha\beta}\neq X^\mu_{\beta\alpha}$ and $Y^\alpha_{\mu\beta}\neq Y^\beta_{\mu\alpha}$;
which holds true for $X^{\prime\mu}_{\alpha\beta}$ and $Y^{\prime\alpha}_{\mu\beta}$ as well.

As in the first iteration earlier on, we now calculate the determinant of the above symplectic two-form,
with views to establishing whether it is singular or not.
Once more, we employ Schur's identity (\ref{eq:detM}), with  $M=\Omega$ in (\ref{eq:omegafinal}), $M_1$ and $M_4$ as in (\ref{eq:idents}) and
\begin{align}
\label{eq:M23def}
\displaystyle
M_2=\left(
\begin{array}{ccccccc}
-\overline{\partial}^\mu_\alpha \mathcal{T}^{00}_\beta &\quad X^\mu_{\alpha\beta}\vspace*{0.2cm}\\
\delta^0_\mu\delta^\alpha_\beta &\quad -Y^\alpha_{\mu\beta} 
\end{array}
\right)\delta^{d-1}(x^i-{x^\prime}^{\, i}), \qquad M_3=\left(
\begin{array}{ccccccc}
\overline{\partial}^\nu_\beta\mathcal{T}^{00}_\alpha &\quad -\delta^0_\nu\delta^\beta_\alpha  \vspace*{0.2cm}\\
-X^{\prime\nu}_{\beta\alpha} &\quad Y^{\prime\beta}_{\nu\alpha} 
\end{array}
\right)\delta^{d-1}(x^i-{x^\prime}^{\, i}).
\end{align}
As an intermediate step, we note that
\begin{align}
\label{eq:intdetres}
\displaystyle
\left(M_3M_1M_2\right)_{\alpha\beta}=\left(
\begin{array}{ccccccc}
0 &\quad Z_{\alpha\beta}  \vspace*{0.2cm}\\
-Z^\prime_{\beta\alpha}  &\quad -X^{\prime\mu}_{\gamma\alpha}Y^\gamma_{\mu\beta}+Y^{\prime\gamma}_{\mu\alpha}X^\mu_{\gamma\beta}
\end{array}
\right)\delta^{d-1}(x^i-{x^\prime}^{\, i}),
\end{align}
where the vanishing components are a direct consequence of the classical consistency conditions (\ref{eq:ourclasscond})
and where we have introduced
\begin{align}
\label{eq:defZab}
Z_{\alpha\beta}:= \left(\overline{\partial}^\mu_\gamma \mathcal{T}^{00}_\alpha\right)Y^\gamma_{\mu\beta}-X^0_{\alpha\beta}.
\end{align}
As explained below (\ref{eq:XYdef}), the primed analogue $Z^\prime_{\alpha\beta}$ stands for $Z_{\alpha\beta}$ under the replacements $\partial_i\leftrightarrow (-)\partial_i^\prime$.
From (\ref{eq:intdetres}), it readily follows that the determinant of $\Omega$ in (\ref{eq:omegafinal}), which we denote by $\varrho$ henceforth, is not zero in general:
\begin{align}
\label{eq:varrhodet}
\varrho=-\textrm{det}\left[(Z^\prime\cdot Z) \,\delta^{d-1}(x^i-{x^\prime}^{\, i})\right]\neq 0.
\end{align}
From (\ref{eq:intdetres}), it is clear that this is a direct consequence of
\begin{align}
\label{eq:quantconscondpart}
Z_{\alpha\beta} \neq 0.
\end{align}
The fact that the above determinant $\varrho$ does {\it not} vanish signals the closure of the symplectic quantization iterative method.

Upon recalling our discussion below (\ref{eq:detOmega1}), a crucial observation follows:
\begin{align}
\label{eq:quantconscond}
\varrho\overset{!}{\neq}0 \implies Z_{\alpha\beta}\overset{!}{\neq} 0.
\end{align}
This is an essential self-consistency condition for the targeted family of massive electrodynamics theories (\ref{eq:ourLag}).
Indeed, if $\varrho=0$, then more than $2N$ constraints would be present.
These can be determined in a third iteration of the symplectic quantization procedure and would over-constrain the theories,
which would no longer enjoy the same constraint algebraic structure of $N$ copies of Proca electrodynamics\footnote{At this point,
the attentive and expert reader may well develop an educated (yet unfounded) suspicion.
Namely, that perhaps $\varrho=0$ is possible, as long as each and every of the additional constraints that follow
are functionally dependent on the already found $2N$ constraints.
However, this is not possible in the targeted theories.
The reason is that a closure of the iterative procedure through functional dependence of the constraints
implies the presence of a (gauge) symmetry.
Clearly, our considered massive electrodynamics theories explicitly break the $U(1)^N$ gauge invariance of $N$-field massless electrodynamics theories
and thus enjoy no symmetry at all. 
Therefore, the iterative algorithm cannot close in such a manner for these theories;
for them, $\varrho=0$ is necessary.
We refer the interested reader to~\cite{ErrastiDiez:2020dux} for further details. \label{indepftnt}}.
We therefore name (\ref{eq:quantconscond}) as {\it quantum consistency conditions} for (\ref{eq:ourLag}).
This complements the classical consistency condition in (\ref{eq:ourclasscond}).
Remarkably and unlike (\ref{eq:ourclasscond}), the new conditions (\ref{eq:quantconscond}) apply to both single and multiple GP field settings.
We regard the unveiling of the quantum consistency conditions as another important result in this paper,
which will be elaborated upon in section \ref{sec:newcons}.

\subsection{Output: the partition function}
\label{sec:output}

The above non-singular symplectic two-form is central to symplectic quantization.
Indeed, the commutation relations between the generalized coordinates (\ref{eq:GCfinal}) are given by
\begin{align}
\label{eq:commrels}
\{Q^n,Q^m\}=\left(\Omega_{mn}\right)^{-1},
\end{align}
with the right-hand side denoting the inverse of $\Omega$ in (\ref{eq:omegafinal}).
It is convenient to make two observations at this point.
First, it is easy to deduce that, in our case, (\ref{eq:commrels}) is {\it not} of the standard canonical form.
This is because in the symplectic piece $(\theta\cdot \dot{Q})$ of our second iterated Lagrangian density
--- where $\theta$ and $Q$ are given by (\ref{eq:thetafinal}) and (\ref{eq:GCfinal}), respectively ---
the set $(\theta,Q)$ is not formed by independent fields: recall (\ref{eq:P0cons}) and (\ref{eq:newconsexpl}).
By construction~\cite{Faddeev:1988qp}, it is guaranteed that there exists a Darboux transformation that brings $(\theta,Q)$ to a canonical set of variables,
whose commutation relations will then be of the standard canonical form.
In general, finding the said Darboux transformation is tedious, if not difficult as well.
Its calculation is a pivotal point in~\cite{Barcelos1,Barcelos}
and finds in~\cite{Prescod-Weinstein:2014lua} what could well be the most complicated worked out example available to date.
In our persistent aim for a quantization without tears,
amenable to extrapolation to more cumbersome Lagrangian densities and aligned with the very essence of the employed method~\cite{Jackiw:1993in},
we omit the determination of such a Darboux transformation.
Second, as a direct consequence of our first observation,
the explicit computation of the inverse matrix in (\ref{eq:commrels}) is operationally lengthy and prone to error.
In fact, it can become quite a mathematical feat to do so, depending on the theory under consideration.
We therefore refrain from its calculation and instead will promptly follow~\cite{Liao,Toms:2015lza},
which will lead to the path integral formulation of the partition function for the theories of our interest (\ref{eq:ourLag}).
For completeness, we note that yet another way around this technical complication was put forward in~\cite{Prescod-Weinstein:2014lua},
which proposes a quantization methodology that markedly departs from the symplectic prescription \`a la Faddeev and Jackiw.

As just anticipated and adhering to~\cite{Liao,Toms:2015lza}, our prior analysis readily yields the sought partition function~\cite{Liao}: 
\begin{align}
\label{eq:partfunct}
Z=\int d\sigma \, \textrm{exp}\left(i\int_{\mathcal{M}}d^dx\, \mathcal{L}\right).
\end{align}
Here, the Lagrangian density $\mathcal{L}$ is of the FJ form (\ref{eq:FJLagform}),
with $\widehat{\mathcal{L}}$, $Q$ and $\theta$ as in (\ref{eq:ourLagprime}), (\ref{eq:GCfinal}) and (\ref{eq:thetafinal}), respectively.
The measure is
\begin{align}
\label{eq:pathintmeasure}
d\sigma=J \Big(\prod_{\mu,\alpha} \left[dA^\alpha_\mu\right] \Big)
\Big(\prod_{\nu,\beta} \left[dp_\beta^\nu \right]\Big)
\Big(\prod_{\gamma} \left[d\lambda^\gamma\right]\Big)
\Big(\prod_{\delta} \left[d\widetilde{\lambda}^\delta\right]\Big),
\end{align}
where $J$ stands for the Jacobian of the aforementioned Darboux transformation.
It is the main result of~\cite{Toms:2015lza} to prove the identification
\begin{align}
\label{eq:starToms}
J=\varrho^{1/2},
\end{align}
with $\varrho$ as in (\ref{eq:varrhodet}) for the theories of our present interest.
The transcendence of (\ref{eq:starToms}) is clear, given our above observation that obtaining the Darboux transformation is generically complicated:
it fully specifies the path integral measure in terms of the central object of the symplectic quantization method
--- the (possibly iterated) non-singular symplectic two-form $\Omega$ ---
in a computationally simple manner.
Therefore and recalling Schwartz's appreciation that ``if you have an exact closed-form expression for $Z$ for a particular theory, you have solved it completely''~\cite{QFTbook},
we have now concluded the symplectic quantization of (\ref{eq:ourLag}) in the path integral formulation\footnote{As a side remark,
we point out that the authors of~\cite{Liao} built upon their own work in~\cite{MoreLiao}, which seems to be a reference that~\cite{Toms:2015lza} is unaware of.
Here, they introduced the so-called {\it equivalently extended Lagrangian}, which does not contain the Jacobian $J$,
as a means to resolve the ambiguity in their prescribed measure for those cases where the Darboux transformation is such that $J\neq 1$.
We find this unillustrated proposal rather obscure and unnecessarily involved and therefore favor the neat resolution of~\cite{Toms:2015lza}.}.

\subsection{Examples}
\label{sec:2simplelimits}

With the main goal of neatly illustrating our above analysis, we proceed to examine two  simple, massive extensions of QED.
We will first contemplate the well-known Proca electrodynamics case and explicitly ensure we reproduce its familiar results.
We then use the developed approach to quantize a single-field GP scenario,
where the mass of the GP field is realized through a derivative self-interaction term. 

\vspace*{0.5cm}

\hspace*{-0.75cm}
\textbf{Proca electrodynamics.}\\
We begin by considering the renowned Lagrangian density
\begin{align}
\label{eq:ProcaLag}
\mathcal{L}_{\textrm{P}}=-\frac{1}{4}A_{\mu\nu}A^{\mu\nu}-\frac{1}{2}m^2A_\mu A^\mu, \qquad m^2\in\mathbb{R}_{>0},
\end{align}
dating back to~\cite{Proca1,Proca2} and subjected to symplectic quantization in e.g.~\cite{Prescod-Weinstein:2014lua,Toms:2015lza,Ramos,Pimentel:2015mza}.
The theoretical appeal of the above singular theory is largely due to the fact that it is the simplest field theory with only second class constraints.
As such, over the years it has been recurrently used in diverse contexts as a representative of the subtleties this class of theories displays during quantization;
for instance, see~\cite{GreiRei,Kim:1996gk,Zamani:2008rt,Silenko:2017qgo,Scham,Park:2019pob}.

Proca electrodynamics is a subcase of our targeted family of theories.
It is obtained by considering the single field limit $N=1$ in (\ref{eq:ourLag}), along with the choices
$c=0=\mathcal{T}^{\mu\nu}$ and in the absence of external sources $J^\mu=0$.

It is immediate to see that the first iterated symplectic two-form (\ref{eq:ouromega1}) has a zero determinant in this case.
Therefore, the classical consistency condition (\ref{eq:ourclasscond}) is automatically satisfied.
The second iterated symplectic two-form (\ref{eq:omegafinal}) always has a non-zero determinant $\varrho$, with
\begin{align}
\label{eq:Procadet}
\varrho^{1/2}=\textrm{det}\left[m^2 \,\delta^{d-1}(x^i-{x^\prime}^{\, i})\right].
\end{align}
Consequently, the quantum consistency conditions (\ref{eq:quantconscond}) are also automatically satisfied.

The partition function of Proca electrodynamics in the symplectic quantization is of the form in (\ref{eq:partfunct}),
where the path integral measure is
\begin{align}
\label{eq:Procadsigma}
d\sigma= \varrho^{1/2} \Big(\prod_{\mu} \left[dA_\mu\right] \Big)
\Big(\prod_{\nu} \left[dp^\nu \right]\Big)
\left[d\lambda\right] \big[d\widetilde{\lambda}\big],
\end{align}
with $\varrho^{1/2}$ as in (\ref{eq:Procadet}), and where the Lagrangian density $\mathcal{L}$ therein is explicitly given by
\begin{align}
\label{eq:ProcacalL}
\mathcal{L}=p^\mu\dot{A}_\mu+p^0\dot{\lambda}+\left(\partial_i p^i-m^2A^0\right)\dot{\widetilde{\lambda}}
-\frac{1}{2}p^ip_i-p^i\partial_i A_0 -\frac{1}{4}A_{ij}A^{ij}-\frac{1}{2}m^2A_\mu A^\mu.
\end{align}
Our above (limiting) result is in agreement with the relevant literature.
We restate that this can be most easily verified by direct comparison to~\cite{Toms:2015lza}.

\vspace*{0.5cm}

\hspace*{-0.75cm}
\textbf{A simple GP electrodynamics.}\\
We proceed to consider the Lagrangian density
\begin{align}
\label{eq:easyGPLag}
\mathcal{L}_{\textrm{GP1}}=-\frac{1}{4}A_{\mu\nu}A^{\mu\nu}+f\partial_\mu A^\mu, \qquad f=f(A_\mu),
\end{align}
which is a subcase of the original GP proposal in~\cite{Tasinato:2014eka,Heisenberg:2014rta}.
There exist preliminary results regarding the quantum behavior of (single field) GP theories~\cite{Charmchi:2015ggf,Amado:2016ugk,deRham:2018qqo,Heisenberg:2020jtr},
also in a curved background~\cite{Panda:2021cbf}.
All of these works are concerned with tree-level and one-loop observables.
However, to our knowledge, no complete and rigorous quantization scheme had been proposed for GP theories prior to this paper.

The above (\ref{eq:easyGPLag}) follows from the single field limit $N=1$ of (\ref{eq:ourLag}), with
\begin{align}
\label{eq:choicesforex2}
m^2=0=c, \qquad \mathcal{T}^{\mu\nu}=f \eta^{\mu\nu},
\end{align}
for no external sources: $J^\mu=0$.

As for Proca electrodynamics earlier on, the first iterated symplectic two-form (\ref{eq:ouromega1}) has a zero determinant.
This is because the classical consistency condition (\ref{eq:ourclasscond}) does not restrict single-GP theories.
Minor algebraic effort yields the following determinant $\varrho$ for the second iterated symplectic two-form (\ref{eq:omegafinal}):
\begin{align}
\label{eq:GProcadet}
\varrho^{1/2}=\textrm{det}\left[\mathcal{F}\,\delta^{d-1}(x^i-{x^\prime}^{\, i})\right], \qquad
\mathcal{F}=(\overline{\partial}^if)\left(\overline{\partial}_if-\partial_i\right)
-(\partial_i A^i)\overline{\partial}^0\overline{\partial}^0f+(p_i+\partial_iA_0)\overline{\partial}^0\overline{\partial}^if-\partial_i\overline{\partial}^i f
\end{align}
which is non-zero for any $f$ that is genuinely a function of the GP field $A_\mu$.
However, the quantum consistency conditions rule out the classical possibility that $f$ be a constant (of suitable length dimension $-2$).
For the simple case here studied, choosing $f$ to be a constant in (\ref{eq:easyGPLag}) renders the mass-like derivative self-interaction into a boundary term,
a case that is obviously of no interest from the very onset.
Therefore, the quantum consistency conditions (\ref{eq:quantconscond}) are also automatically satisfied in our second simple example.

Symplectic quantization gives rise to partition function of (\ref{eq:easyGPLag}) in the form (\ref{eq:partfunct}),
where the measure is as in (\ref{eq:Procadsigma}), with  $\varrho^{1/2}$ given by (\ref{eq:GProcadet}),
and where
\begin{align}
\label{eq:GProcacalL}
\displaystyle
\begin{array}{lllll}
\mathcal{L}=&p^\mu\dot{A}_\mu+\big(p^0+f\big)\dot{\lambda}
+\left[\partial_i p^i +\left(p_i+\partial_i A_0\right)\overline{\partial}^if+\left(\partial_i A^i\right)\overline{\partial}^0f\right]\dot{\widetilde{\lambda}} \vspace*{0.2cm}\\
&-\frac{1}{2}p^ip_i-p^i\partial_i A_0 -\frac{1}{4}A_{ij}A^{ij}+f\partial_i A^i.
\end{array}
\end{align}

\section{Quantum consistency conditions}
\label{sec:newcons}

The above symplectic quantization of (\ref{eq:ourLag}) has revealed two insights.
On the one hand, the necessarily singular character of the first iterated symplectic two-form (\ref{eq:ouromega1})
implies the (already known) classical consistency conditions (\ref{eq:ourclasscond}).
On the other hand, the necessarily non-singular character of the second iterated symplectic two-form (\ref{eq:omegafinal})
implies the (newly found) quantum consistency conditions (\ref{eq:quantconscond}).
If any given theory within (\ref{eq:ourLag}) fails to fulfill (\ref{eq:ourclasscond}), then this theory is ill-defined at the classical level.
Specifically, it would be prone to Ostrogradski instabilities~\cite{Ostrogradsky:1850fid}.
If any given theory within (\ref{eq:ourLag}) fulfills (\ref{eq:ourclasscond}) but not (\ref{eq:quantconscond}),
then this theory does not admit quantization.
Namely, such a theory must be exclusively viewed as a classical effective field theory (EFT);
it cannot be employed as a quantum EFT.

The violation of the quantum consistency conditions (\ref{eq:quantconscond}) 
should {\it not} be interpreted as an anomaly, i.e.~the quantum breaking of a classical symmetry.
This is because, in any (multi-)GP electrodynamics theory, the gauge symmetry is explicitly broken already at the classical level.
Moreover, the violation of (\ref{eq:quantconscond}) should {\it not} be regarded as related to a symmetry enhancement,
wherein multi-GP (partially) restores the $U(1)^N$ gauge symmetry of $N$ copies of Maxwell electrodynamics or its massless non-linear extensions.
(Multi-)GP explicitly breaks the gauge symmetry, regardless of whether the quantum consistency conditions are satisfied or not.
An easy way to see this is as follows.
Consider the example (\ref{eq:easyGPLag}).
This Lagrangian density enjoys a $U(1)$ gauge symmetry when either $f=0$ or $\partial_\mu A^\mu=0$.
The quantum consistency conditions for this theory imply that $\mathcal{F}$ in (\ref{eq:GProcadet}) cannot vanish. 
Since $f=0$, $\partial_\mu A^\mu=0$ and $\mathcal{F}=0$ are functionally independent formulae,
we readily deduce that there exists no relation between the violation of the quantum consistency conditions and the restoration of a gauge symmetry in the theory.

In full generality, the class of multi-GP electrodynamics theories in section~\ref{sec:review}
can be reasonably expected to reproduce the above described structure.
Namely, the 
imperative singularity of their first iterated symplectic two-form
presumably implies the classical consistency conditions (\ref{eq:SecCons}), while the essential
non-singularity of their second iterated symplectic two-form
{presumably} implies the suitable generalization of the quantum consistency conditions (\ref{eq:quantconscond}) to
\begin{align}
\label{eq:quantconscondgen}
P \overset{!}{\neq} 0 {\implies \tilde{\mathcal{Z}}_{\alpha\beta}\overset{!}{\neq}0},
\end{align}
with $P$ the determinant of the second iterated symplectic two-form
{and $\tilde{\mathcal{Z}}_{\alpha\beta}$ the appropriate
extension of $Z_{\alpha\beta}$ in (\ref{eq:defZab})}. 
We emphasize that (\ref{eq:quantconscondgen}) affects a large class of theories.
For instance, it restricts in an unprecedented manner any GP electrodynamics theory,
wherein the mass of the GP field is realized {\it exclusively} through derivative self-interactions.
This means considering a single-field $N=1$ and setting $\mathcal{L}_{(0)}=0$ with $\mathcal{L}_{(n\geq 1)}\neq0$ in (\ref{eq:LagInt4}),
for one or more such $n\geq 1$.
In this case, (\ref{eq:quantconscondgen}) rules out the classically consistent possibility of having constant $\mathcal{T}$ objects for $\mathcal{L}_{(n\geq 2)}$,
since  (\ref{eq:quantconscondgen}) necessarily involves at least one derivative with respect to the GP field. 
Thus, we conclude that, for the general multi-GP electrodynamics theories reviewed in section \ref{sec:review},
the $\mathcal{T}$'s are non-trivially constrained by (\ref{eq:SecCons}),
as well as by our newly found quantum consistency conditions (\ref{eq:quantconscondgen}).

\section{Conclusions and outlook}
\label{sec:out}

In this work, we have carried out the symplectic quantization of the family of multi-field Generalized-Proca (GP) electrodynamics theories in (\ref{eq:ourLag}).
Specifically, we have determined the partition function (\ref{eq:partfunct}).
As a by-product, we have obtained an independent derivation of the classical consistency conditions (\ref{eq:ourclasscond}) that apply to these theories.
Moreover, we have unveiled a necessary additional set of restrictions for (multi-)GP theories in the quantum regime,
which we call quantum consistency conditions (\ref{eq:quantconscond}).
Remarkably, these affect both single- and multi-field scenarios and imply that (most but) not all
generalizations of massive electrodynamics considered here can be quantized. 

It is possible  that our newly found quantum consistency conditions, even when generically fulfilled for a given Lagrangian, are dynamically violated.
For the family of theories (\ref{eq:ourLag}), this would mean that there exists one or more points in the moduli space
for which (\ref{eq:quantconscond}) does not hold true.
In the second example considered in section~\ref{sec:2simplelimits}, 
this is realized when the generalized coordinates $A_\mu$ and $p^i$ take on-shell values that result in the vanishing of
$\mathcal{F}$ in (\ref{eq:GProcadet}).
This type of singularities in the second iterated symplectic two-form would imply the existence of further constraints in the theory,
which, if functionally independent, would lead to a reduction of the local number of physical modes.
We have explicitly checked that such reduction of the local degrees of freedom indeed takes place in the example  (\ref{eq:easyGPLag}),
for the particular choice $f=-A_\mu A^\mu/2$.
Phenomena like shock wave propagation and birefringence are then expected to occur. 

Indeed, similar degenerate behavior is theoretically well-known to happen in the massless sector: in the family of theories known as non-linear electrodynamics (NLE) --- recall table \ref{table:EMtable}.
For instance, shock waves have been studied in the particular NLE cases of Born electrodynamics~\cite{Minzetal,Kadlecova:2021bmp},
and of Euler-Heisenberg electrodynamics~\cite{Kadlecova:2018vty}, as well as generically in the Plebanski formulation of the full NLE family~\cite{Escobar:2020zcm}
(see also references therein).
Born-Infeld electrodynamics constitutes the only sensible exception within NLE: this theory displays no shock waves and no birefringence~\cite{Boillat}.

{\color{blue}As already noted in 
section~\ref{sec:intro},}  massless scenarios in NLE are currently pending experimental verification.
Our work suggests that analogue massive settings in (multi-)GP
should be phenomenologically studied and confronted with the outcome of the relevant future experiments,
such as PVLAS~\cite{Ejlli:2020yhk} and LUXE~\cite{Abramowicz:2021zja}.
A particularly appealing question to be addressed is the examination 
of whether the class of multi-GP electrodynamics theories contains a subset which, like Born-Infeld,
completely avoids degenerate behavior.


\vspace*{0.5cm}

\noindent
{\bf Acknowledgments:}
We are grateful to Dieter~L\"{u}st and Julio~A.~M\'endez-Zavaleta for their careful reading of an earlier version of this paper.
Their incisive questions and insightful comments have notoriously improved the presentation of our results.
The work of VED is funded by the Deutsche Forschungsgemeinschaft (DFG, German Research Foundation) under Germany's Excellence Strategy -- No. EXC-2094 -- 390783311.


\end{document}